\documentclass[aps,pral,showpacs,superscriptaddress,amssymb,twocolumn]{revtex4-2}
\usepackage{graphicx}
\usepackage{appendix} 
\usepackage{hyperref}
\usepackage{bm,amsmath}
\usepackage{dcolumn}

\usepackage{soul}
\usepackage[x11names]{xcolor}

\newif\ifARXIV
\newif\ifUSEBBL

\ARXIVtrue
\USEBBLtrue

\def\Z2{ {\mathbb Z}_{2} }

\newcommand{\BRA}[1] {\langle  #1 |}

\newcommand{\KET}[1] {| #1 \rangle}

\def\SPINSQUEEZING{KitagawaUedaPRA1993,HaldSorensenPRL1999,KuzmichMandelPRL2000,WangSanders2003,JinKimPRL2007,MaWangPhysRep2011,DegenReinhardRMP2017,EsteveGrossNat2008,PezzeSmerziRMP2018,ChaiLaoPRL2020,XinChapmanPRXQ2022,XinBarriosPRL2023}

\begin{document}

\rightline{\tt  }

\vspace{0.2in}

\title[]{Emergent Non-Markovian Nonlinear Qubit From Collective Spin Interactions
% in the Kitagawa-Ueda model
}

\author{Gregory T. Carroll}
%\email{mgeller@uga.edu}
\affiliation{Department of Physics and Astronomy, University of Georgia, Athens, Georgia 30602, USA}
\affiliation{Center for Simulational Physics, University of Georgia, Athens, Georgia 30602, USA}

\author{Michael R. Geller}
%\email{mgeller@uga.edu}
\affiliation{Department of Physics and Astronomy, University of Georgia, Athens, Georgia 30602, USA}
\affiliation{Center for Simulational Physics, University of Georgia, Athens, Georgia 30602, USA}

\author{Andr\'e Erpenbeck}
%\email{mgeller@uga.edu}
\affiliation{Department of Physics and Astronomy, University of Georgia, Athens, Georgia 30602, USA}
\affiliation{Center for Simulational Physics, University of Georgia, Athens, Georgia 30602, USA}

\begin{abstract}
Open-system descriptions are typically introduced by coupling a quantum system to an external environment. Here we show that a closed interacting many-body system can itself generate a controlled non-Markovian quantum channel acting on a reduced nonlinear qubit through finite-size corrections to a nonlinear mean-field limit. 
We demonstrate this using the Kitagawa-Ueda one-axis twisting model,
$H=\chi J_z^2$, a paradigmatic model of collective spin dynamics, spin squeezing, and two-component Bose-Einstein condensates.
Although the large-$N$ regime of this model has been extensively studied, the conventional fixed-$\chi$ scaling does not yield a nontrivial dynamical large-$N$ limit. 
In this paper, we investigate a complementary large-$N$ formulation obtained from the double limit $N\rightarrow\infty$ and $\chi\rightarrow O(g/N)$, where $g$ is a coupling constant. 
We derive the leading finite-$N$ corrections to this limit and show that they correspond to an emergent non-Markovian dephasing process, producing a Gaussian decay of the Bloch-vector coherence with characteristic timescale $t_\varphi\geq\sqrt{N}/(2g)$.
Exact finite-$N$ calculations demonstrate that this effective open-system description becomes quantitatively accurate for systems containing on the order of one hundred qubits. The resulting framework provides a microscopic realization of non-Markovian dephasing generated intrinsically by a closed many-body system and enables efficient simulation of collective quantum dynamics beyond unitary mean-field theory.
These results link the long-studied phenomenon of phase diffusion in atomic ensembles and Bose-Einstein condensates to the growing effort to characterize non-Markovian, beyond-Lindblad noise in quantum computing hardware, providing a rare case in which such a noise channel is derived from microscopic dynamics rather than fit phenomenologically.

\end{abstract}

\maketitle
\setcounter{tocdepth}{2}
\clearpage
%\tableofcontents

\section{Introduction}

Understanding how effective open-system dynamics emerge from closed interacting quantum systems is a central challenge in quantum many-body physics and an important problem for scalable quantum technologies. While decoherence is traditionally modeled by coupling a quantum system to an external environment, interacting many-body systems contain internal degrees of freedom that can themselves act as effective environments for their local constituents. As quantum processors continue to increase in size, interactions among qubits constitute an intrinsic source of correlated dynamics that cannot, in general, be described by independent Markovian noise models \cite{Rivas_Quantum_2014, Breuer_Colloquium_2016, Agarwal_Modelling_2024, Nakamura_Qubit_2024, Zou_Spatially_2024}. Developing controlled microscopic descriptions of these emergent reduced-system dynamics is therefore important both for understanding collective quantum behavior and for accurately modeling large-scale quantum devices.

Spin models with all-to-all Ising interactions play a central role in quantum information, condensed matter physics, and atomic physics. The canonical form is
\begin{eqnarray}
H_{\rm CG} = \frac{\chi}{2}\sum_{i<j}\sigma_i^z\sigma_j^z,
\qquad
\chi\in\mathbb{R},
\label{cg model}
\end{eqnarray}
where $\sigma^\mu$ $(\mu\in\{x,y,z\})$ are Pauli matrices. Because the interaction graph is complete, we refer to this as the complete-graph (CG) model. Introducing the collective spin operator
\begin{eqnarray}
J_\mu=\sum_{i=1}^{N}\frac{\sigma_i^\mu}{2},
\label{angular momentum from spins}
\end{eqnarray}
the Hamiltonian can be written (up to a constant shift) as the Kitagawa-Ueda (KU) one-axis twisting model \cite{KitagawaUedaPRA1993},
\begin{eqnarray}
H_{\rm CG}=H_{\rm KU}-\frac{\chi N}{4},
\qquad
H_{\rm KU}=\chi J_z^2.
\label{ku model}
\end{eqnarray}
The permutation symmetry of this model greatly simplifies the many-body problem while retaining highly nontrivial collective dynamics, making the KU model a paradigmatic example of interacting quantum many-body physics.

A major application of all-to-all interactions is the generation of spin squeezing and multipartite entanglement
\cite{\SPINSQUEEZING,230915353,240219429}. The resulting spin-squeezed states exhibit reduced quantum fluctuations and have become an important resource for quantum metrology. Experimentally, one-axis twisting has been realized in a variety of platforms, including cold atomic ensembles \cite{\SPINSQUEEZING} and cavity-QED systems \cite{230915353,240219429}, where collective coupling to a common mode generates effective $J_z^2$ interactions. These developments have established the KU model as one of the canonical settings for studying collective quantum dynamics.

Despite the extensive literature on the KU model, most studies of its large-$N$ behavior consider the conventional limit of fixed interaction strength $\chi$ \cite{KitagawaUedaPRA1993, MaWangPhysRep2011, PezzeSmerziRMP2018}. In this limit, however, the nonlinear dynamics become increasingly rapid with increasing system size and do not possess a well-defined dynamical thermodynamic limit. In this work we instead investigate the complementary scaling
$\chi=g/N$, which preserves finite nonlinear dynamics as $N$ increases and leads to a well-defined large-$N$ limit described by a nonlinear mean-field qubit. This scaling provides a controlled connection between collective many-body dynamics and nonlinear single-qubit evolution.

Building on this formulation, we derive the leading finite-size corrections to the nonlinear mean-field limit. Our central analytical result is that these corrections take the form of a Gaussian decay of the transverse qubit coherence with characteristic timescale
$t_\varphi\propto\sqrt{N}$. We further show that this reduced dynamics is exactly reproduced by a time-local non-Markovian dephasing master equation, providing a microscopic realization of non-Markovian open-system dynamics generated entirely by the intrinsic dynamics of a closed interacting many-body system rather than by coupling to an external environment \cite{Breuer_Colloquium_2016,Rivas_Quantum_2014}. Finally, by comparing the effective description with the exact finite-$N$ solution of the KU model, we determine the regime in which the nonlinear open-qubit description becomes quantitatively accurate and demonstrate that it provides an efficient framework for describing collective quantum dynamics beyond unitary mean-field theory.

\section{Emergent nonlinear qubit in the large-$N$ limit} \label{sec:large-N}

\subsection{Finite $N$ magnetization} 

The time-dependent nonequilibrium magnetization in the KU model can be calculated exactly, starting in any $N$-spin permutation-symmetric product 
state 
\begin{eqnarray}
\KET{\Psi_{\! N}} =  \KET{\psi}^{\! \otimes N}
\! \!  ,  \ \ 
\KET{\psi}  = \alpha  \KET{0}  + \beta \KET{1} , \ \ 
| \alpha |^2  \! + \! | \beta |^2 = 1. \ \ \ 
\label{product state} 
\end{eqnarray}
Here $ \KET{\psi} $ is a single spin (qubit) state. 
Let  $J_{+} = J_x + i J_y$ and
use $[J_z , J_{+} ] = J_{+}$ to obtain
$ U^\dagger J_{+} U = J_{+}  \, e^{2 i  \chi t  J_z} e^{i \chi t }$, 
where $U= e^{ -iH_{\rm KU} t }= e^{-i \chi  t J_z^2}.$ Then from  (\ref{angular momentum from spins}) we have
\begin{eqnarray}
 \langle J_{+} \rangle 
&=& \BRA{\Psi_{N}} J_{+}   
e^{2 i \chi t J_z} e^{i \chi t} 
\KET{\Psi_{N}} , \nonumber \\
&=&  \frac{N}{2} e^{i \chi t} 
 \BRA{\psi} (\sigma^{x}  + i \sigma^{y} ) 
 e^{i \chi t \sigma^{z}  }   \KET{\psi} \,
  \big[ \BRA{\psi}  e^{i \chi t \sigma^{z}  }   \KET{\psi} \big]^{N-1} ,  \nonumber \\
  &=& \frac{N}{2} (x_0 + i y_0) 
\big[  \cos(\chi t) + i z_0  \sin(\chi t) 
\big]^{N-1} .
\label{exact jplus expression}
\end{eqnarray}
Here $x_0, y_0, z_0$ are the Bloch coordinates of the initial state $\KET{\psi}$. 
In addition, 
\begin{eqnarray}
 \langle   J_{z} \rangle = \frac{N}{2} z_0 ,
\end{eqnarray}
because $J_{z} $ commutes with the Hamiltonian.
Thus, we have obtained the magnetization 
 $\langle {\vec  J} \rangle$ for finite $N$.

\subsection{Mean field magnetization} 

We expect the $N \rightarrow \infty $ limits of the $H_{\rm CG}$ and $H_{\rm KU}$ models to lead to a single-qubit nonlinear mean field Hamiltonian
\begin{eqnarray}
H_{\rm eff} = g \langle \sigma^z \rangle  \sigma^z , \ \ g \in {\mathbb R},
\label{heff}
\end{eqnarray}
called the ($z$ axis) torsion model
\cite{MielnikJMP80,150706334,211105977,240219429}.
To see this, 
expand  $H_{\rm CG}$ in fluctuations 
 $\sigma_i^z - \langle \sigma^z \rangle $
(assuming permutation symmetry).
Neglecting quadratic fluctuations leads to
\begin{eqnarray}
H_{\rm CG} &=& \frac{\chi}{4} \sum_{i \neq j} \sigma_i^z \sigma_j^z
\approx 
 \frac{\chi}{4} \sum_{i \neq j} 
 \bigg[  \langle \sigma^z \rangle^2
 + 2 \langle \sigma^z \rangle   
\big( \sigma_i^z - \langle \sigma^z \rangle \! 
\big) \bigg] \nonumber  \\
&=& N \bigg[ 
g \langle \sigma^z \rangle   \sigma^z
- \frac{g}{2} \langle \sigma^z \rangle^2 \bigg] , 
\ \ g := \frac{\chi(N-1)}{2}.
\label{mft}
\end{eqnarray}
So in the mean field  limit we obtain 
$N$ identical copies of (\ref{heff}), 
plus an energy shift.

The mean field magnetization is obtained
from (\ref{angular momentum from spins}),
which leads to
\begin{eqnarray}
\langle {\vec J}  \rangle = \frac{N}{2} \langle {\vec \sigma}  \rangle = \frac{N}{2} {\vec r} , \end{eqnarray}
where
\begin{eqnarray}
{\vec r} := \langle {\vec \sigma}  \rangle = {\rm tr} ( \rho {\vec \sigma} )
\label{bloch vector definition}
\end{eqnarray}
is the Bloch vector of the mean field qubit
and $\rho$ is the reduced density matrix 
(many-body density matrix traced over all spins but one). Therefore in the $N \rightarrow \infty$ limit
we have 
\begin{eqnarray}
\frac{d \rho}{dt} = -i [ H_{\rm eff} , \rho ]  , \ \ 
\rho(0) = |\psi \rangle \langle \psi |,
\label{master equation pure}
\end{eqnarray}
where $ \KET{\psi} $ is the initial single-qubit state
[see (\ref{product state})],
but $O(N^{-1})$ corrections instead produce a mixed state due to phase decoherence.
This mean-field torsion dynamics, and the phase decoherence appearing at finite $N$, are the collective-spin analog of quantum phase diffusion studied in two-mode and split Bose-Einstein condensates \cite{Lewenstein_Quantum_1996, Javanainen_Phase_1997, Castin_Relative_1997}.

 \subsection{$N \rightarrow \infty$ limit}
 
Next we take the double limit 
\begin{eqnarray}
N \rightarrow \infty \ \ {\rm and} \ \  \chi \rightarrow  \frac{2g}{N \! - \! 1} ,
\label{double limit}
\end{eqnarray}
with $g$ constant. The factor of two in
(\ref{double limit}) follows conventional notation for $g$. 
The factor $(N-1)^{-1}$ can be replaced by $N^{-1}$ in 
the large $N$ limit, but (\ref{double limit}) is preferred because it leads to a relation between $\chi$ and $g$ that is also correct to order $N^{-1}$
(ignoring this would lead to an unphysical frequency shift at order $N^{-1}$). Both features also follow directly from the definition of $g$ in 
(\ref{mft}).

To understand the dynamics of
$ \langle {\vec J} \rangle $ in the large $N$ limit,
set  $\chi \rightarrow 2g (N \! - \! 1 )^{-1}$  and write (\ref{exact jplus expression}) as
\begin{eqnarray}
 \langle J_{+} \rangle 
=  \frac{N}{2} (x_0 + i y_0) \, f(t) ,
\label{exact jplus  alt}
\end{eqnarray}
where
\begin{eqnarray}
f(t) := \big[ \! \cos( \omega t) + i z_0 \sin(\omega t) \big]^{N-1} ,\ \  f(0) = 1.
\label{f definition}
\end{eqnarray}
Here  $\omega =  2g(N-1)^{-1}$.
In Sec.~\ref{maclaurin series} we analyze the large $N$ behavior of $f(t)$. In terms of this function,
the angular momentum components evolves as
\begin{eqnarray}
&&  \frac{  \langle   J_{x} \rangle }{N/2} 
= x_0 \,  {\rm Re}(f) - y_0 \, {\rm Im}(f) , \\
&&  \frac{  \langle   J_{y} \rangle }{N/2} 
= x_0 \,  {\rm Im}(f) + y_0 \, {\rm Re}(f) ,  \\
&& \frac{  \langle   J_{z} \rangle }{N/2} = z_0 .
\end{eqnarray}
The Bloch vector (\ref{bloch vector definition})
evolves as
\begin{eqnarray}
\begin{pmatrix} x \\ y \\  z  \end{pmatrix}
= T
\begin{pmatrix}  x_0 \\ y_0 \\  z_0 \end{pmatrix} \! \! ,\ \  T := 
\begin{pmatrix}
{\rm Re} (f)  &  -{\rm Im}  (f)   & 0   \\   {\rm Im}  (f)  & {\rm Re} (f)  & 0  \\ 0 & 0 & 1
\end{pmatrix} \! \! .
\end{eqnarray}
Here $T$ is the time-evolution operator for the Bloch vector $ {\vec r} = (x,y,z);$  it is nonlinear due to its dependence on $ z_0$. 

In the next section  we show  that (also see \cite{Geller2026fromspinsqueezingto})
\begin{eqnarray}
\lim_{N \rightarrow \infty}  f(t) =  e^{2 i g z_0 t} .
\label{f infinity}
\end{eqnarray}
Then in this limit we have
\begin{equation}
\begin{pmatrix} x \\ y \\ z \end{pmatrix}
=
\begin{pmatrix} 
\cos (2gz_0t) & -\sin (2gz_0t) & 0  \\
\sin (2gz_0t) & \cos (2gz_0t) & 0  \\ 
0  & 0  & 1  \\ 
\end{pmatrix}
\begin{pmatrix} x_0 \\ y_0 \\ z_0 \end{pmatrix} \!  .
\label{t infinity}
\end{equation}
The matrix implements $z$-axis torsion, namely a $z$ rotation of the Bloch vector with frequency $2g z_0$.
   
\subsection{Maclaurin series coefficients as moments}
\label{maclaurin series}

Our main objective is to go beyond
(\ref{f infinity}) and (\ref{t infinity}) and calculate the $1/N$ corrections to $f(t)$ and $T$.
Because $f(t)$ is analytic at $t=0$ we can express it as a Maclaurin series in time:
\begin{eqnarray}
f(t) = 1 + \sum_{k=1}^\infty \frac{t^k}{k!} f^{(k)}(0) , 
\ \ f^{(k)}(t) := \frac{d^k \! f}{dt^k} .
\end{eqnarray}
Here we show how to calculate the Maclaurin series coefficients $f^{(k)}(0)$ to order $1/N$ by relating them to the $k^{\rm th}$ moment of a random variable in a certain binomial distribution.
To proceed, write (\ref{f definition}) as
\begin{eqnarray}
f(t) &=&  \bigg[ 
\bigg( \frac{1+z_0}{2} \bigg)  \, e^{i \omega t} +
\bigg( \frac{1-z_0}{2} \bigg)  \, e^{-i \omega t} \bigg]^{N-1}  \\
&=& \sum_{m=0}^{n} \binom{n}{m} \,
p^m  \, (1-p)^{n-m}  \, e^{ i \omega 
(2m-n) t}  , \ \ 
\end{eqnarray}
where
\begin{eqnarray}
n = N-1 \ \ {\rm and} \ \ 
p = \bigg( \frac{1+z_0}{2} \bigg) .
\end{eqnarray}
Then
\begin{eqnarray}
 f^{(k)}(0) = \sum_{m=0}^{n} \binom{n}{m} \,
p^m  \, (1-p)^{n-m} 
[i \omega (2m-n) ]^k \!  \ \ \ \\
 =  (2 i \omega)^k \ 
 \sum_{m=0}^{n} \binom{n}{m} \,
p^m  \, (1-p)^ {n-m} \
 \big( m - a \big)^k  \! , \ \ \
 \label{raw expression for coefficient}
\end{eqnarray}
where $a = n/2$.  We note that
 (\ref{raw expression for coefficient})
can be interpreted as
[$( 2 i \omega  )^{ k}$ times]  
a moment of the
binomial distribution: 
\begin{eqnarray}
f^{(k)}(0) &=& 
 \bigg(\frac{4 i g }{n} \bigg)^{\! k} \, 
\sum_{m=0}^{n}  
{\rm Pr}(m) \, \big(m - a \big)^k  \\
&=& \bigg(\frac{4 i g }{n} \bigg)^{\! k} \, 
  \big\langle (X - a)^k \big\rangle .
\label{general expression for coefficient}
\end{eqnarray}
Here 
\begin{eqnarray}
X = x_1 + x_2 + \cdots + x_n, \ \ x_i \in \{ 0,1\}
\end{eqnarray}
is the sum of independent and identically distributed  Boolean random variables, 
the bracket in (\ref{general expression for coefficient}) denotes expectation, and
\begin{eqnarray}
{\rm Pr}(X\! \!  = \!  \! m)  = \binom{n}{m} \, p^m (1-p)^{n-m} , 
\end{eqnarray}
is the probability of observing $m \, 1{\rm s }$ 
among  $n$ random bits, 
with the single-bit probability
\begin{eqnarray}
 {\rm Pr}(x_i\! \! = \! \!1) = p = \bigg( \frac{1 + z_0}{2} \bigg).
\end{eqnarray}
Recall that
\begin{eqnarray}
\langle X \rangle &=& \sum_{m=0}^{n}  
{\rm Pr}(m) \, m = n p ,  \\
\langle X^2 \rangle &=& \sum_{m=0}^{n}  
{\rm Pr}(m) \, m^2 = n p (1-p) + n^2 p^2,
\end{eqnarray}
and
\begin{eqnarray}
{\rm Var}(X) =
\langle X^2 \rangle - \langle X \rangle^2  
=  n p (1-p)  .
\end{eqnarray}

\begin{widetext}

Having set up the problem, 
use $ X-a =  \frac{n z_0}{2} + (X - \langle X \rangle) $  and expand the moment as 
\begin{eqnarray}
(X - a)^k \!
&=& \! \bigg( \frac{n z_0}{2}  \bigg)^k
 \bigg[ 1 + k \bigg( \frac{2}{nz_0} \bigg)
 (X - \langle X \rangle)
 + \frac{k(k-1)}{2}\bigg( \frac{2}{nz_0} \bigg)^2
  (X - \langle X \rangle)^2 +  \cdots  \bigg] .
 \end{eqnarray}
 Then
 \begin{eqnarray}
\langle (X - a)^k \rangle
  =  \bigg( \frac{nz_0}{2}  \bigg)^k
 \bigg[ 1  + \frac{k(k-1)}{2}\bigg( \frac{2}{nz_0} \bigg)^2
n p (1-p) +  \cdots  \bigg] ,
 \end{eqnarray}
and
 \begin{eqnarray}
f^{(k)}(0) &=&
(2 i z_0 g)^k
\bigg[ 1 + \frac{k(k-1)(1-z_0^2)}{2 N z_0^2}
 \bigg] +  O(N^{-2})  ,
\label{large n expression for coefficient}
\end{eqnarray}
which is the final result for the large $N$ Maclaurin series coefficients.

\end{widetext}

Then for any fixed time $t$ and large $N$ we have
\begin{eqnarray}
f(t) &=& 1 + \sum_{k=1}^\infty 
\frac{ \lambda^k }{k!} 
\bigg[ 1 +
k(k-1) \bigg(  \frac{1-z_0^2}{2 N z_0^2} \bigg)
 \bigg] \nonumber \\ &+& O(N^{-2})  , 
\end{eqnarray}
where  $\lambda =  2 i z_0 g t.$ 
Then we find
\begin{eqnarray}
f(t) &=& e^{\lambda }  \bigg[ 1 + 
\lambda^2 \bigg( \frac{1-z_0^2}{ 2 N z_0^2} \bigg)  
 \bigg] +  O(N^{-2}) \\
&=& e^{2 i z_0 g t } 
\bigg[ 1 -  \frac{2  (1-z_0^2) g^2 t^2}{N} 
\bigg] +  O(N^{-2}) \\
&=&
e^{ - \frac{ 2 (1-z_0^2) g^2 t^2}{N} }
 e^{2 i z_0 g t  }   \big[ 1 + O(N^{-2}) \big] ,
\end{eqnarray}
leading to the final result
\begin{eqnarray}
f(t) = e^{ - \frac{1}{2} (t/ t_\varphi  \! )^2 }
 e^{2 i z_0 g t }   \big[ 1 + O(N^{-2}) \big] ,
  \label{large n expression for f}
\end{eqnarray}
where
\begin{eqnarray}
t_\varphi 
= \sqrt{ \frac{N}{4 (1-z_0^2) g^2} } .
\label{decoherence timescale}
\end{eqnarray}
% Equation (\ref{large n expression for f}) is our main result. 
Equation (\ref{large n expression for f}) is the central result of this work, it shows that the finite-size corrections to collective unitary dynamics take the form of an effective Gaussian decay with timescale
(\ref{decoherence timescale}). 
% This decay is caused by the small entanglement generated when $N$ is large but finite.  
This reduced single-qubit decoherence arises from entanglement generated between one spin and the remaining collective degrees of freedom.
An interesting prediction is that the depolarization time   
(\ref{decoherence timescale}) is
 proportional to $\sqrt{N}$, and it diverges 
(with a square-root singularity) as the classical states $z_0 \rightarrow \pm 1$ are approached. 
The bound
\begin{eqnarray}
t_\varphi \ge  \frac{\sqrt{N}}{2g}
\label{short time entanglement time}
\end{eqnarray}
provides a worst-case estimate of 
$ t_\varphi $.

\section{Emergence and validation of the effective non-Markovian channel} \label{sec:numerics}
\subsection{Validity of the Large-$N$ Approximation}\label{sec:validity}

To assess the predictability of the large $N$ expression in Eq.~(\ref{large n expression for f}) and its decoherence timescale, we determine the system sizes for which the large-$N$ description becomes quantitatively accurate.
To this end, we evaluate the exact finite $N$ expression defined in Eq.~(\ref{f definition}) for the representative parameters $\omega = 2g/(N-1)$, $g=1$, $\theta=0.4$, and $\phi=0.6$, and compare the resulting dynamics to the corresponding large $N$ prediction.

\begin{figure}[t]
    \centering
    \includegraphics[width=\columnwidth]{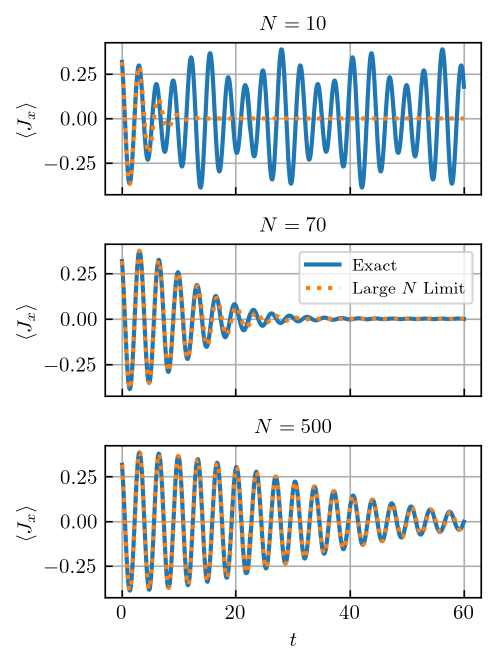}\\
    \vspace*{-11.25cm}
    \raggedright(a)\\
    \vspace*{3.15cm}
    (b)\\
    \vspace*{+3.15cm}
    (c)
    \vspace*{+3.75cm}
    \caption{Normalized expectation value of the $x$ component of the angular momentum operator, $\langle J_x \rangle /(N/2)$, as a function of time for different system sizes, $N=10$, $70$, and $500$. The solid lines show the exact finite $N$ result obtained from Eq.~(\ref{f definition}), while the dashed lines correspond to the large $N$ approximation given in Eq.~(\ref{large n expression for f}).}
    \label{fig:JxVsT}
\end{figure}

Figure~(\ref{fig:JxVsT}) shows the normalized expectation value of the angular momentum operator in the $x$ direction, $J_x$, for several representative qubit numbers, ranging from small systems, where the large $N$ prediction is expected to break down, to large systems, where the large $N$ approximation should become accurate. This behavior is clearly reflected in our results: for larger system sizes (middle and bottom panels), the large $N$ expression for $f(t)$ closely reproduces the exact dynamics. In contrast, for small system sizes (top panel), we observe pronounced deviations between the two descriptions. While the large-$N$ limit predicts a Gaussian decay, the exact finite size result is dominated by short time revivals that inhibit a meaningful decay process. This leads to qualitatively different physical behavior in the two descriptions for small $N$.

Beyond the underlying dynamics, we assess the validity of the large $N$ prediction for the decoherence timescale $t_\varphi$, an important quantity for quantum systems and quantum hardware applications. To this end, we compare the analytical expression in Eq.~(\ref{decoherence timescale}) with the decoherence timescale extracted directly from the exact finite-$N$ function $f(t)$. The extracted value is obtained by averaging over $N_{\mathrm{timesteps}}$ sampled times,
\begin{equation}
t_\varphi^{\mathrm{extracted}}
=
\frac{1}{N_{\mathrm{timesteps}}}
\sum_t
\left[
\frac{-2}{t^2}
\log
|f(t)|
\right]^{-1/2},
\end{equation}
where we use $N_{\mathrm{timesteps}}=1000$ and data evaluated on a time window ranging from $1$ to $t_{\max}=5t_\varphi$. 

\begin{figure}[t]
    \raggedright (a)\\
    \centering
    \includegraphics[width=\columnwidth]{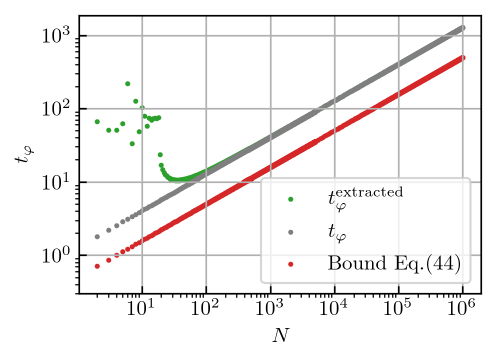}\\
    \raggedright (b)\\
    \includegraphics[width=\columnwidth]{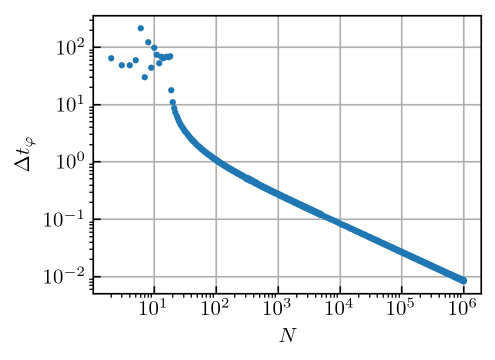}
    \caption{
    Decoherence timescale as a function of system size $N$. 
    The analytical prediction $t_\varphi$ from Eq.~(\ref{decoherence timescale}) is compared with the decoherence timescale extracted from the exact finite $N$ numerical data, $t_\varphi^{\mathrm{extracted}}$.
    (a) Extracted decoherence timescale together with the large $N$ prediction from Eq.~(\ref{decoherence timescale}) and the corresponding lower bound for the decoherence timescale.
    (b) Absolute deviation between the analytical prediction and the numerically extracted value, $\Delta t_\varphi = |t_\varphi - t_\varphi^{\mathrm{extracted}}|$.}
    \label{fig:t_phiVals}
\end{figure}

Figure~\ref{fig:t_phiVals}(a) shows the extracted decoherence timescales as a function of system size $N$. For small $N$, the extracted decoherence time is highly volatile and does not exhibit a clear trend. This behavior is a consequence of the strongly oscillatory finite-$N$ dynamics, for which the assumption of a smooth decay process is not applicable, rendering the notion of a decoherence timescale ill defined. Around $N \sim 20$, the extracted values begin to show a more systematic dependence on $N$, indicating that the Gaussian decay form becomes increasingly well defined, although noticeable deviations from the large $N$ prediction remain.
For larger systems, $N \gtrsim 100$, the extracted decoherence time $t_\varphi^{\mathrm{extracted}}$ becomes nearly indistinguishable from the analytical prediction on the scale of the plot, providing an estimate for the system size at which the large $N$ description becomes reliable.

To quantify the convergence of the decoherence time, Fig.~\ref{fig:t_phiVals}(b) shows the absolute deviation between the analytical prediction and the extracted value,
$\Delta t_\varphi = \left| t_\varphi - t_\varphi^{\mathrm{extracted}} \right|$.
This deviation as a function of $N$ further corroborates our analysis and  provides a direct connection between a desired accuracy and the required number of qubits. In particular, for $N \gtrsim 100$, the data indicate systematic power-law convergence toward the large $N$ limit, consistent with the scaling in Eq.~(\ref{decoherence timescale}).

\subsection{Effective Non-Markovian Dynamics}

The Gaussian coherence decay obtained in the large-$N$ limit naturally suggests an effective open-system description of the reduced qubit dynamics. In contrast to the exponential decay generated by time-independent Markovian Lindblad equations \cite{Lindblad1976, Gorini_Completely_1976, NielsenChuang2010}, the Gaussian envelope derived in Eq.~(\ref{large n expression for f}) requires a time-dependent decay rate and is naturally described by a time-local non-Markovian master equation \cite{Breuer_Colloquium_2016,Rivas_Quantum_2014}. Motivated by this microscopic result, we construct an effective nonlinear master equation and show that it quantitatively reproduces the large-$N$ dynamics.

Before turning to the specific form of the emergent channel, it is worth asking whether its operator structure is dictated by symmetry alone or by the particular microscopic mechanism generating it. To address this, we compare two natural candidate channels, one that respects the axial symmetry of the underlying interaction, and one that does not.
The first corresponds to pure dephasing,
\begin{eqnarray}
\frac{d \rho}{dt} =
-i \left[ H_{\rm eff}, \rho \right]
+
\frac{\gamma(t)}{2}
\left(
\sigma^z \rho \sigma^z-\rho
\right),
\label{dephasing master equation}
\end{eqnarray}
while the second corresponds to isotropic depolarization,
\begin{eqnarray}
\frac{d \rho}{dt} =
-i \left[ H_{\rm eff}, \rho \right]
+
\frac{\gamma(t)}{4}
\sum_{\mu\in\{x,y,z\}}
\left(
\sigma^\mu \rho \sigma^\mu-\rho
\right).
\label{depolarization master equation}
\end{eqnarray}
In both cases, the nonlinear mean-field Hamiltonian is
\begin{eqnarray}
H_{\rm eff}=g\langle\sigma^z\rangle\sigma^z,
\end{eqnarray}
and the time-dependent rate is chosen as
\begin{eqnarray}
\gamma(t)=\frac{t}{t_\varphi^2},
\end{eqnarray}
such that the resulting coherence decay reproduces the Gaussian envelope
$e^{-\frac{1}{2}(t/t_\varphi)^2}$
of the large-$N$ solution. The explicit time dependence of $\gamma(t)$ is the defining feature of the non-Markovian effective description introduced below.

\begin{figure*}[t]
    \centering
    \includegraphics[width=\textwidth]{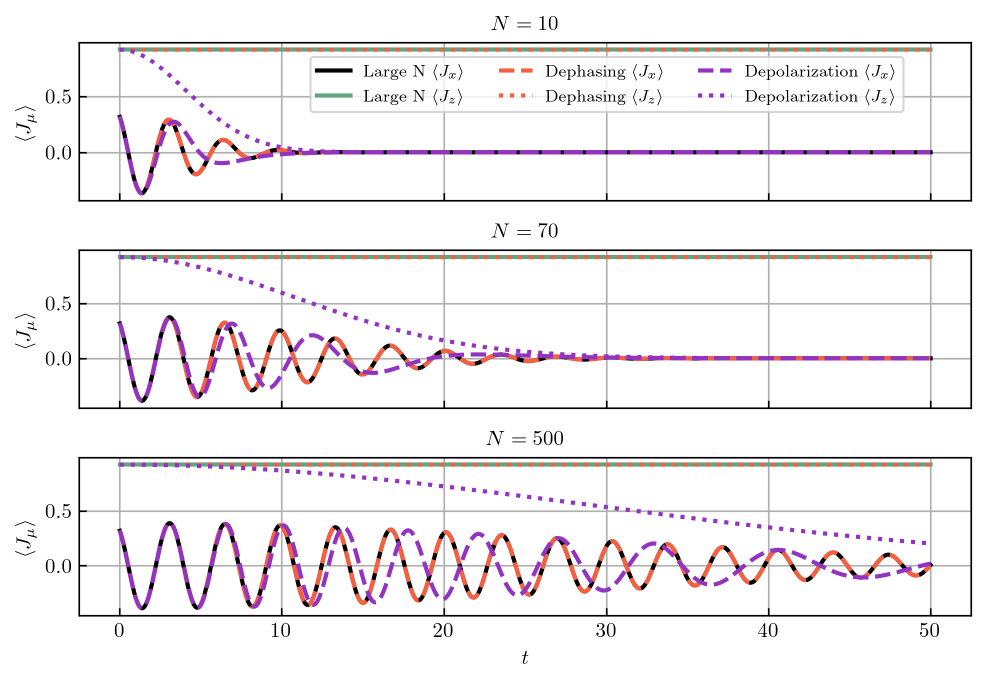}
    \caption{
        Normalized collective-spin expectation values
        $\langle J_\mu\rangle/(N/2)$ for
        $\mu\in\{x,z\}$ as functions of time for
        $N=10$, $70$, and $500$.
        The solid black curves show the large-$N$
        result from Eq.~(\ref{large n expression for f}).
        The orange and purple curves correspond to numerical solutions of the effective non-Markovian master equations with dephasing
        (Eq.~\ref{dephasing master equation})
        and isotropic depolarization
        (Eq.~\ref{depolarization master equation}),
        respectively.
    }
    \label{fig:NoiseModels}
\end{figure*}

Figure~\ref{fig:NoiseModels} compares the dynamics obtained from the effective master equations with the large-$N$ solution of Eq.~(\ref{large n expression for f}). Time evolution was performed using an explicit Euler integration method with timestep $\Delta t=10^{-5}$ \cite{Press2007Numerical}. The numerical propagation conserves the density-matrix trace,
$\mathrm{Tr}(\rho)\approx1$,
to better than $10^{-10}$ over the full simulation interval. %, obviating the need for a more refined propagation algorithm. 
All remaining parameters are identical to those used in Sec.~\ref{sec:validity}.

The pure dephasing model reproduces the large-$N$ dynamics with essentially perfect agreement for all system sizes considered. In particular, it captures the Gaussian decay of the transverse component $\langle J_x\rangle$ while preserving the conserved longitudinal polarization $\langle J_z\rangle$. This behavior reflects the structure of the underlying one-axis twisting Hamiltonian, where the decay of coherence originates from phase dispersion rather than relaxation.

In contrast, the isotropic depolarization model fails to reproduce the collective-spin dynamics beyond short times. While it captures the initial decay of transverse coherence, it additionally suppresses the longitudinal component $\langle J_z\rangle$, leading to relaxation absent in the KU model. The comparison therefore identifies pure dephasing as the appropriate effective channel generated by finite-size corrections to the nonlinear mean-field dynamics.
Moreover, this comparison illustrates a more general point: the appropriate effective open-system description of a microscopically generated channel is not fixed by generic symmetry considerations alone, but is determined by the specific structure of the underlying many-body interaction. A phenomenological Lindblad ansatz, such as the isotropic depolarizing channel considered here, can be a poor guide to the correct effective dynamics even when it is qualitatively reasonable, underscoring the value of deriving the channel microscopically rather than assuming its form.

The structure of this emergent channel becomes explicit by writing the corresponding Bloch-vector equations. For the dephasing model,
\begin{eqnarray}
\frac{d\vec r}{dt}
=
\vec u\times\vec r
-
\gamma(t)
\left[
\vec r-(\vec r\cdot\vec e_z)\vec e_z
\right],
\qquad
\vec u=2gz\,\vec e_z ,
\label{nonlinear master equation bloch vector}
\end{eqnarray}
where $\vec r=(x,y,z)$ denotes the Bloch vector. The nonlinear Hamiltonian generates precession around the $z$ axis, while the dissipative contribution acts only on the transverse components. Consequently, the Bloch vector acquires a Gaussian decay of phase coherence in the $x$--$y$ plane while its longitudinal component remains constant, reproducing the characteristic collapse of the KU dynamics.
For comparison, the isotropic depolarization model gives
\(
\frac{d\vec r}{dt}
=
\vec u\times\vec r
-
\gamma(t)\vec r ,
\label{nonlinear master equation bloch vector depolar}
\)
which causes a simultaneous decay of all components of the Bloch vector and therefore does not represent the effective dynamics generated by the KU model.

The emergent dephasing process has a fundamentally different origin from conventional environmental decoherence. In the strict $N\rightarrow\infty$ limit, the mean-field description becomes exact and the dynamics are purely unitary. The effective decoherence discussed here arises from finite-size corrections and reflects the small amount of entanglement generated between the collective qubit and the remaining degrees of freedom. For permutation-symmetric systems, this is consistent with the suppression of multipartite entanglement in the thermodynamic limit \cite{CoffmanPRA00,OsborneVerstraetePRL2006,YangPLA2006,Geller2026fromspinsqueezingto}.
From a semiclassical perspective, Gaussian coherence decay is typically associated with slowly varying fluctuations in an effective noise field. In the KU model, the corresponding fluctuations originate intrinsically from finite-size fluctuations of the collective spin. 
This is closely analogous to the collapse of coherence observed for number-state superpositions evolving under a number-conserving nonlinear Kerr-type interaction, most directly realized in the collapse of the matter-wave field of a Bose-Einstein condensate in an optical lattice \cite{Greiner_Collapse_2002}.
As the system size increases, the characteristic timescale grows as
$t_\varphi\propto\sqrt{N}$,
and the effective dephasing vanishes in the infinite-size limit, recovering the purely unitary nonlinear qubit dynamics.

\section{Conclusions}
\label{sec:conclusion}

In this work, we have shown that finite-size corrections to a closed collective quantum system can give rise to an effective non-Markovian open-system description. Using the Kitagawa-Ueda one-axis twisting model as a paradigmatic example, we introduced a large-$N$ formulation based on the scaling $\chi=2g/(N-1)$, which yields a well-defined nonlinear mean-field qubit limit. We then derived the leading finite-$N$ corrections to this limit and demonstrated that the resulting reduced qubit dynamics are accurately described by a non-Markovian dephasing process with Gaussian coherence decay and characteristic timescale $t_\varphi \geq \sqrt{N}/(2g)$.

This emergent open-system description provides a different perspective on finite-size effects in collective quantum dynamics. Rather than introducing decoherence phenomenologically through an external environment, the effective dephasing channel arises from the intrinsic many-body degrees of freedom of the interacting ensemble. Importantly, the resulting dynamics extend beyond the standard Lindblad framework commonly used in quantum information and quantum computing \cite{Lindblad1976, Gorini_Completely_1976, NielsenChuang2010}, where Markovian noise models are typically assumed. The non-Markovian master equation derived here provides a controlled example of how microscopic many-body dynamics can generate memory effects in effective quantum channels.

Beyond its conceptual implications, the nonlinear open-qubit description provides a practical route for simulating collective quantum dynamics in regimes where direct many-body approaches become challenging. Our numerical results show that the effective description accurately captures the finite-$N$ dynamics already for ensembles containing on the order of one hundred qubits, while reducing the description from an $N$-spin interacting system to a single effective nonlinear qubit. 
This enables efficient modeling of collective quantum systems at sizes beyond the reach of many conventional simulation approaches used for interacting quantum systems.

The framework developed here suggests several directions for future investigation. Extending this approach to more general collective spin Hamiltonians, spatially structured interactions, and other nonlinear mean-field models may reveal additional examples where finite-size many-body corrections naturally generate effective non-Markovian quantum dynamics. Furthermore, the explicit connection between microscopic collective interactions and emergent reduced-system dynamics may provide a useful framework for characterizing interaction-induced decoherence in quantum devices. In platforms where collective spin interactions are engineered, such as atomic ensembles or quantum processors with tunable interactions, this approach could serve as a tool for quantifying deviations from ideal unitary evolution and identifying signatures of correlated non-Markovian effects.
This connects to a growing body of work characterizing non-Markovian, beyond-Lindblad noise directly on superconducting-qubit hardware, where residual qubit-qubit coupling, for example spectator-induced $ZZ$ crosstalk, generates memory effects that are, in spirit, an engineering-scale analog of the intrinsic many-body coupling studied here: in both settings, a qubit's effective environment is not external but is built from its interaction with the very degrees of freedom it remains entangled with \cite{Agarwal_Modelling_2024, Nakamura_Qubit_2024, Zou_Spatially_2024, Li_Probing_2026}. 
A key distinction is that the channel identified here is derived, rather than fit, directly from a microscopic Hamiltonian. The same strategy of starting from a known interaction graph and computing the reduced dynamics exactly, rather than assuming a generic Lindblad or post-Markovian ansatz, may prove useful for identifying the structure and rate of noise channels in engineered processors, where qubit-qubit couplings are often known but their reduced-system consequences are not. We emphasize that the specific mechanism identified here relies on permutation-symmetric, all-to-all coupling. Extending this microscopic strategy to the more structured, typically local connectivity of superconducting-qubit arrays is a natural and, we believe, promising direction for future work.
More broadly, our results establish a connection between collective quantum dynamics and emergent open-system descriptions, providing a controlled framework for studying quantum evolution beyond unitary mean-field theory.

% physically we need to tune chi to zeor

\acknowledgements

This work was partly supported by the NSF under grant no.~DGE-2152159.
G.T.C.\ and A.E.\ were supported by startup funds from the University of Georgia. ChatGPT (OpenAI) was used to assist with language and text editing of the manuscript. All scientific content, analysis, and conclusions were developed and verified by the author.

\section*{Data Availability}

Data underlying the results presented in this paper are not publicly available at this time. They are available from the authors upon reasonable request.

%\appendix
%
%\section{Recommended parameter values}

\bibliography{bib.bib}

%\ifARXIV
%\else
%\leftline{\large \bf References}
%\fi

%\bibliographystyle{unsrtnat}
%\ifUSEBBL
%\bibliography{MS3.bbl}
%\else
%\bibliography{/Users/mgeller/Dropbox/bibliographies/CM,/Users/mgeller/Dropbox/bibliographies/MATH,/Users/mgeller/Dropbox/bibliographies/QFT,/Users/mgeller/Dropbox/bibliographies/QI,/Users/mgeller/Dropbox/bibliographies/group,/Users/mgeller/Dropbox/bibliographies/books}
%\fi

%%%%%%%%%% Merge SI %%%%%%%%%%%%%%%%%%%
% Prefix an "S" to all equations, figures, tables and reset the counter 
%%%%%%%%%%%%%%%%%%%%.%%%%%%%%%%%%%%

\end{document}

\setcounter{equation}{0}
\setcounter{figure}{0}
\setcounter{table}{0}
\setcounter{page}{1}
\setcounter{section}{0}
\setcounter{secnumdepth}{4}
\makeatletter
\renewcommand{\thesection}{\arabic{section}}
\renewcommand{\theequation}{S\arabic{equation}}
\renewcommand{\thefigure}{S\arabic{figure}}
\renewcommand{\bibnumfmt}[1]{[S#1]}
\clearpage
\onecolumngrid
\begin{center}
\Large{ Supplementary Information for \\ ``Quantum Simulation of Operator Spreading in the Chaotic Ising Model''}
\end{center}

\vspace{1cm}
\twocolumngrid

This document provides additional details.

\end{document}